\documentclass[12pt, draftclsnofoot, onecolumn]{IEEEtran}
\ifCLASSINFOpdf
\else
\fi
\pdfoutput=1
\usepackage{amsmath}
\usepackage{graphicx}
\usepackage{amsthm}
\usepackage{mathrsfs}
\usepackage{caption}
\usepackage{amsfonts}
\usepackage{amsmath}
\usepackage{amsbsy}
\usepackage{amssymb}
\usepackage{etoolbox}
\usepackage{algorithm}
\usepackage{algorithmic}
\usepackage{subfig}
\usepackage{mathrsfs}
\usepackage{multirow}
\usepackage{cite}
\usepackage{diagbox}
\usepackage{float}

\usepackage{bm}
\begin{document}
%
\title{A Learnable Optimization and Regularization Approach to Massive MIMO CSI Feedback}
%
%
%

\author{\IEEEauthorblockN{Zhengyang Hu, Guanzhang Liu, Qi Xie, Jiang Xue,~\IEEEmembership{Senior Member,~IEEE},\\
Deyu Meng,~\IEEEmembership{Member,~IEEE}, and Deniz G\"und\"uz,~\IEEEmembership{Fellow,~IEEE}}
\thanks{Z. Hu, G. Liu, Q. Xie, J. Xue and D. Meng are with School of Mathematics and Statistics, Xi'an Jiaotong University, Xi'an, 710049, China (e-mail: hzyxjtu@stu.xjtu.edu.cn, lgzh97@stu.xjtu.edu.cn, xie.qi@xjtu.edu.cn, x.jiang@xjtu.edu.cn and dymeng@mail.xjtu.edu.cn).}\\
\thanks{D. G\"und\"uz is with the Department of Electrical and Electronic Engineering, Imperial College London,
London SW7 2AZ, U.K. (e-mail: d.gunduz@imperial.ac.uk).}
\thanks{The corresponding author is J. Xue.}
}

\maketitle

\begin{abstract}
Channel state information (CSI) plays a critical role in achieving the potential benefits of massive multiple input multiple output (MIMO) systems. In frequency division duplex (FDD) massive MIMO systems, the base station (BS) relies on sustained and accurate CSI feedback from the users. However, due to the large number of antennas and users being served in massive MIMO systems, feedback overhead can become a bottleneck. In this paper, we propose a model-driven deep learning method for CSI feedback, called learnable optimization and regularization algorithm (LORA). Instead of using $l_1$-norm as the regularization term, a learnable regularization module is introduced in LORA to automatically adapt to the characteristics of CSI. 
We unfold the conventional iterative shrinkage-thresholding algorithm (ISTA) to a neural network and learn both the optimization process and regularization term by end-to-end training. We show that LORA improves the CSI feedback accuracy and speed. Besides, a novel learnable quantization method and the corresponding training scheme are proposed, and it is shown that LORA can operate successfully at different bit rates, providing flexibility in terms of the CSI feedback overhead. Various realistic scenarios are considered to demonstrate the effectiveness and robustness of LORA through numerical simulations.
\end{abstract}

\begin{IEEEkeywords}
Massive MIMO; CSI feedback; model-driven; deep learning; regularization learning.
\end{IEEEkeywords}

%
\IEEEpeerreviewmaketitle

\section{Introduction}
%
%
%
%
\IEEEPARstart {A}{s} a core technology for the sixth generation (6G) of wireless networks, massive multiple input multiple output (MIMO) systems can provide higher data rates and link reliability \cite{6375940}. To realize the benefits provided by massive MIMO, such as beamforming \cite{7084118} and more reliable signal detection \cite{9159940}, accurate channel state information (CSI) at the base station (BS) is necessary in both the time division duplex (TDD) and frequency division duplex (FDD) modes. In the TDD mode, downlink CSI can be obtained directly from uplink CSI under the assumption of perfect channel reciprocity. However, the TDD mode may not work well in time sensitive scenarios, such as live streaming and vehicular communications \cite{9497358}. In the FDD mode, the uplink and downlink use different frequency resources at the same time. However, due to the lack of perfect channel reciprocity in the FDD mode, the user equipments (UEs) need to estimate downlink CSI and feed it back to the BS \cite{7434506}. Nevertheless, the huge feedback overhead due to the large number of antennas at the BS and the large number of users being served can become a significant performance bottleneck. Therefore, a CSI feedback method with low overhead and high accuracy is essential to deliver the promised gains of massive MIMO systems in next generation communication networks.

\par It was shown in \cite{zhou2007experimental} and \cite{1203156} through experiments that shared local scatterers lead to correlations across CSI in spatial and frequency domains with the increasing scale of  antennas in massive MIMO systems. Following 2-dimensional discrete Fourier transformation (2D-DFT), CSI is shown to exhibit approximate sparsity in the angular-delay domain,
which means that CSI can be compressed to reduce the feedback overhead \cite{8284057}.

\par Compressive sensing (CS) based methods can be used to project sparse signals to a low-dimensional space and recover them efficient with theoretical guarantees. The first CS based CSI feedback method for massive MIMO systems was proposed in \cite{6214417}, which considers both 2D-DFT and Karhunen-Loeve Transform (KLT) as sparsifying bases. In \cite{7417036}, the authors used the statistical information about the angle-of-departure (AOD) to develop a basis for sparsity mapping and a weighted $l_1$-norm was proposed for recovery, which achieves a better performance than the DFT basis. Considering orthogonal frequency-division multiplexing (OFDM) systems, a multidimensional CS-based analog CSI feedback method was proposed in \cite{6966062}, which treats the CSI feedback design as a multidimensional matrix compression and recovery problem, and exploited tensor decomposition. However, these methods are limited in general as they cannot identify the best basis, and the projected CSI matrices are often not perfectly sparse, resulting in performance loss. Although some particular priors are shown to reduce the high requirement of sparsity \cite{metzler2016denoising, 6816089, 7417036}, these `manual' designs are done in a case-by-case bias and are still not efficient enough due to the diverse use cases and high performance requirements of future systems.

\par In recent years, data-driven methods, in particular, deep learning (DL), has achieved notable success in a variety of wireless communication applications \cite{8839651}, such as channel estimation \cite{8640815}, signal detection \cite{9159940}, joint source-channel coding \cite{9834372}, decoding \cite{9086486}, and beamforming \cite{9518251}. DL-based methods have also made tremendous strides for CSI feedback. Specifically, we identify two main groups of works. First group includes convolution based methods, where a convolutional neural network (CNN) is trained on channel data to reduce its dimension. In the second group, we consider model-driven methods.

The authors in \cite{8322184} were the first to employ DL method for CSI feedback and proposed a simple CNN auto-encoder architecture for dimensionality reduction, which has been considered as a baseline for most DL-based CSI feedback methods. The encoder and decoder in \cite{8322184} carry out the compression and recovery operations, respectively. Since increasing the receptive field in CNN can extract more information from the input, CsiNet+ in \cite{8972904} considers different convolution kernel dimensions. Inspired by the inception model, the authors in \cite{9149229} designed CRNet, which uses multi-paths and multi-receptive fields in both encoder and decoder to improve the performance. MRFNet proposed in \cite{9495802} shows that the larger number of convolution channels can recover more details of CSI. In \cite{yang2019deep}, projected CSI coefficients are further quantized, and entropy coded to reduce the required rate. A significant improvement was reported with respect to CsiNet \cite{8322184}. This approach was extended to CSI feedback from multiple nearby users in \cite{9296555}, where the correlation among CSI matrices is exploited to achieve better compression efficiency. These methods all benefit from effective CNN design technics. Researchers have also adopted other DL techniques for CSI feedback. The authors in \cite{9169908} proposed a CSI feedback method based on generative adversarial networks (GANs). In \cite{9585309}, CSI feedback is modeled as an image super resolution problem, and SRNet is proposed. To extract time correlation of CSI, CNN-LSTM-A \cite{zhang2020massive} and CsiNet-LSTM \cite{wang2018deep} are proposed, respectively. Although the aforementioned CNN based methods have achieved significant performance improvements compared to their CS-based counterparts, these methods simply treat the channel matrix as a two-dimensional `image' with local correlations, which may limit their performance.

\par Model-driven DL methods exploit our prior knowledge about the particular learning problem. Bringing model-driven and data-driven approaches together, model-driven DL methods not only make the learned model more explainable and predictable \cite{xu2018model}, but also avoid the requirements for accurate and explicit modeling. In \cite{wu2019compressed}, the authors proposed a model-driven DL method to improve the recovery accuracy in CSI feedback, by unfolding a conventional CS algorithm into a neural network (NN) and learning the measurement matrix. Inspired by the transformation matrix design and unfolding, TiLISTA-Joint was proposed in \cite{wang2020learnable}, which not only learns the down-sampling matrix, but also uses sparse auto-encoder to learn the sparse transformation. To further improve the recovery accuracy, the authors exploited the attention mechanism for sparse transformation learning and proposed FISTA-Net in \cite{9663378}.

\par Although the model-driven DL methods have exhibited remarkable success in CSI feedback, current methods are all designed with an $l_1$-norm regularization term, which cannot extract the prior knowledge of CSI in some cases. Actually, how to design a suitable regularization (i.e., data prior) is an enduring problem in machine learning. It is well-known that $l_0$-norm is the optimal regularization term to describe sparsity, but the optimization with $l_0$-norm is untractable. When the measurement matrix satisfies restricted isometry property (RIP) condition, $l_1$-norm is equivalent to $l_0$-norm in terms of sparse signal recovery \cite{7931570}. Besides, the authors in \cite{meng2013robust} utilized a mixture of Gaussian distributions to learn the noise distribution. Although the proposed method in \cite{meng2013robust} does not explicitly formulate a regularization term, the data prior has been learned in the loss function. Due to the strength of DL, the authors proposed a proximal dehaze-net, which learns a haze-related prior to achieve the obvious performance gain in single photo dehazing \cite{9502677}. In \cite{9157358}, the authors proposed the RCDNet to automatically extract the prior from rain images for better deraining.

Inspired by the model-driven methods for CSI feedback and regularization term learning in \cite{meng2013robust, 9502677, 9157358}, in this paper, we propose a joint regularization and optimization method, called learnable optimization and regularization algorithm (LORA). LORA exploits a NN to learn the regularization term for better fitting the characteristics of CSI, and develops an iterative algorithm with learnable parameters to achieve performance gains.


\par The main contributions of this work are summarized as follows:
\begin{itemize}
  \item Existing model-driven DL architectures for CSI feedback all unfold the algorithm derived from an optimization problem with the $l_1$-norm regularization term, which cannot describe the prior of CSI well due to its imperfect sparsity. Instead, LORA treats the regularization term as a learnable function that can be adjusted according to the characteristics of CSI itself. The proposed method results in a novel algorithm of model-driven DL for CSI feedback with significant improvements in the performance.
\end{itemize}


\begin{itemize}
  \item To mitigate the effect of quantization in LORA, we exploit quantization-aware training (QAT) with learnable quantization parameters, such as quantization scale and zero point value. The proposed quantization method eases the performance decay caused by quantization in different bit levels.
\end{itemize}


\begin{itemize}
  \item The numerical results show that LORA has a superior performance than CsiNet+ \cite{8972904}, CRNet \cite{9149229}, TiLISTA-Joint \cite{wang2020learnable} and ISTA-NET in different scenarios based on 3GPP TR 38.901 \cite{7060514}. Moreover, the performance with channel estimation error and complexity comparisons are provided. We also carry out ablation studies to explore the effects of different modules of LORA on the final performance.
\end{itemize}


\par The rest of this work is organized as follows. Section \uppercase\expandafter{\romannumeral2} describes the massive MIMO system, CSI feedback procedure and channel model. We re-formulate the CSI feedback problem and present the basic algorithm in Section \uppercase\expandafter{\romannumeral3}. In Section \uppercase\expandafter{\romannumeral4}, we present LORA with insights in detail. Numerical results and analyses are provided in Section \uppercase\expandafter{\romannumeral5} to demonstrate the superiority of LORA compared to the existing CSI feedback schemes. Finally, the paper is concluded in Section \uppercase\expandafter{\romannumeral6}.

\textbf{Notations:} Throughout the paper, bold uppercase letters, bold lowercase letters and non-bold letters are used to denote matrices, vectors and scalars, respectively. $\|\cdot\|_2$ is the Euclidean norm. $|\cdot|$ stands for element-wise absolute value. $(\cdot)^T$ and $(\cdot)^H$ are transpose and conjugate transpose, respectively. The real and complex number fields are $\mathbb{R}$ and $\mathbb{C}$, respectively. The expectation operation is represented by $\mathbb{E}\{\cdot\}$.

\section{System Model}

\subsection{Massive MIMO system and CSI feedback}

\begin{figure}[htbp]
\renewcommand{\thefigure}{1}
\centering
\includegraphics[width=3in]{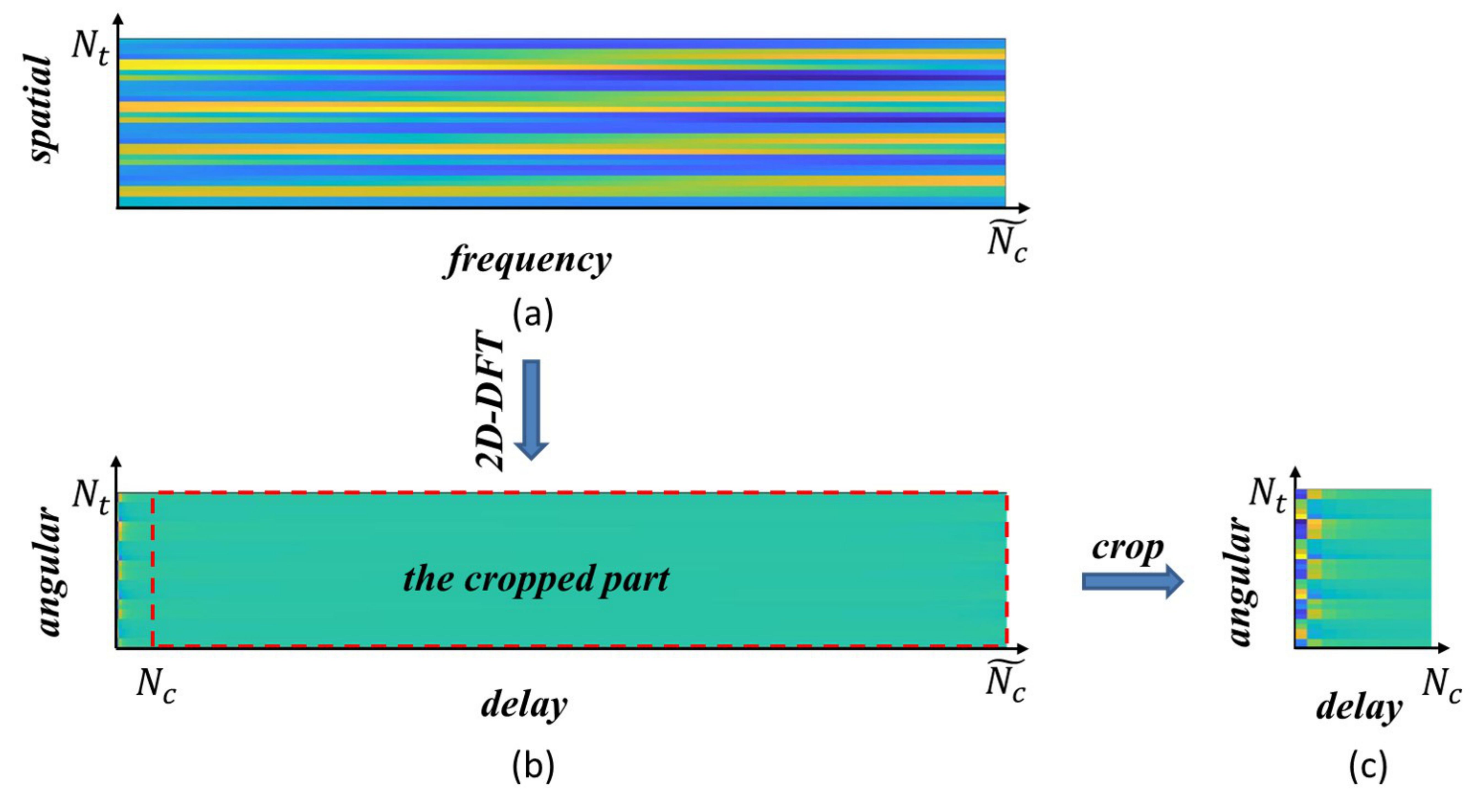}
\caption{The real part of the complete channel matrix in (a) spatial-frequency domain, the real part of the complete (b) and the cropped (c) channel matrices in angular-delay domain.}
\end{figure}

We considers the downlink of a single-cell massive MIMO OFDM system in the FDD mode, where a BS equipped with $N_t \gg 1$ antennas serves single-antenna user equipment (UE) \cite{8322184} over $\tilde{N_c}$ subcarriers. The received signal at the $n$-th subcarrier $(n=1,\ldots,\tilde{N_c})$ in the frequency domain can be expressed as
\begin{equation} \label{e1}
y_n = \tilde{\bm{h}}_n^H\bm{v}_nx_n+z_n,
\end{equation}
where $\tilde{\bm{h}}_n \in \mathbb{C}^{N_t \times 1}$, $\bm{v}_n \in \mathbb{C}^{N_t \times 1}$ and $x_n \in \mathbb{C}$ are the downlink channel vector, corresponding precoding vector and the modulated transmitted signal, respectively, while $z_n \sim \mathcal{CN}(0,1)$ denotes the random additive Gaussian noise. The detailed channel model will be introduced in Section II-B. In the FDD mode, the downlink channel vector $\tilde{\bm{h}}_n$ has to be estimated at the UE and sent back to the BS. The overall downlink CSI matrix can be expressed as $\tilde{\bm{H}} = [\tilde{\bm{h}}_1, \tilde{\bm{h}}_2, \cdots, \tilde{\bm{h}}_{\tilde{N_c}}]$, which consists of $2N_t\tilde{N_c}$ real numbers.
We can transform $\tilde{\bm{H}}$ from spatial-frequency domain to angular-delay domain by 2D-discrete Fourier transformation (DFT) as
\begin{equation} \label{e3}
\bm{H} = \bm{F}_d \tilde{\bm{H}} \bm{F}_a,
\end{equation}
where $\bm{F}_d \in \mathbb{C}^{{N}_t \times {N}_t}$ and $ \bm{F}_a \in \mathbb{C}^{ \tilde{N_c} \times \tilde{N_c}}$ are DFT matrices. As mentioned, $\bm{H}$ has approximate sparsity in angular-delay domain. Moreover, only the first few rows of $\bm{H}$ have significant values because the delay between multipath components typically  lies within a limited period in the delay domain. So, we preserve only the first $N_c<\tilde{N}_c$ rows and remove the rest. An example illustrating the real part of the complete and cropped channel matrix in  angular-delay domain, and the complete channel matrix in spatial-frequency domain are shown in Fig. 1.

\begin{figure*}[t]
    \renewcommand{\thefigure}{2}
    \centering{
    \includegraphics[width=3.7in]{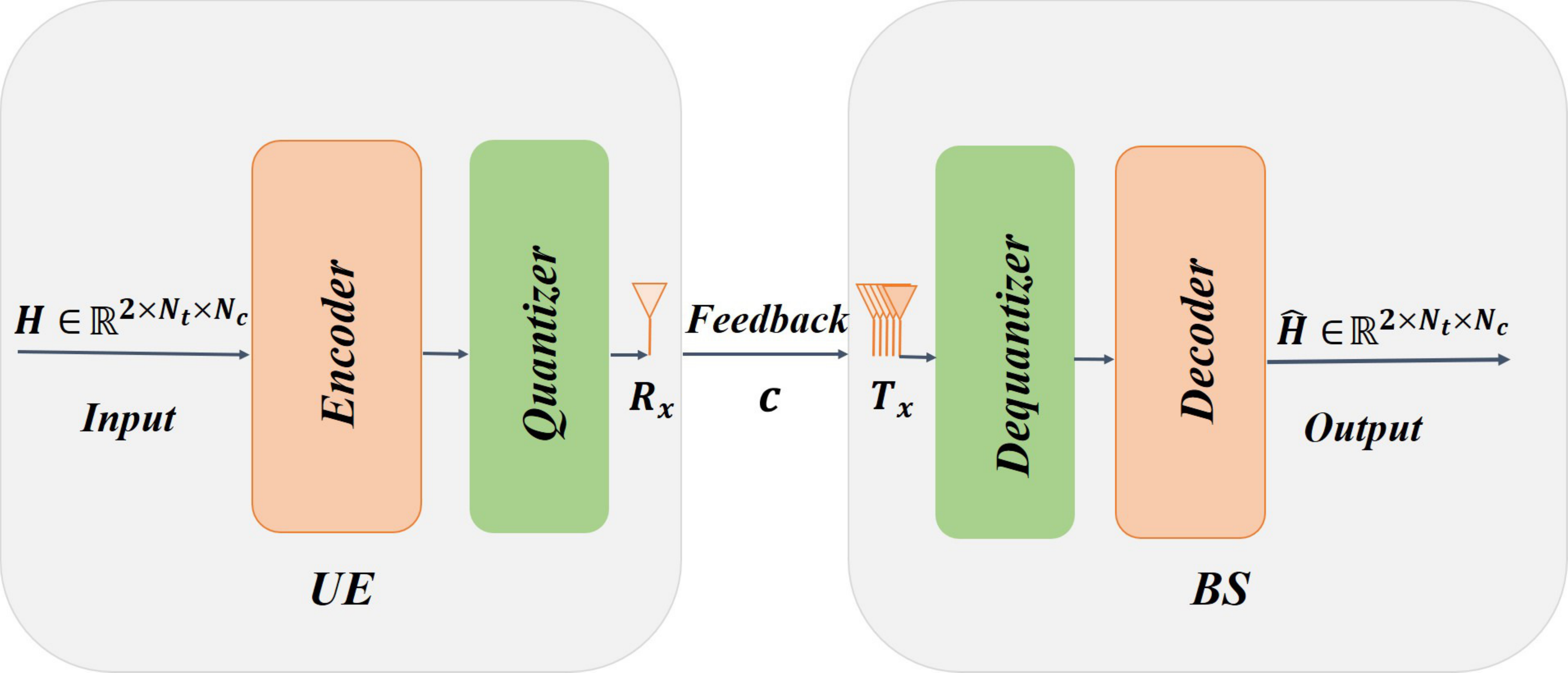}}
    \caption{CSI feedback procedure.}
\end{figure*}

In this case, $\bm{H}$ is still used to denote the truncated CSI. Although the required number of feedback parameters has been reduced from $2N_t \times \tilde{N_c}$ to $2N_t \times N_c$, the feedback overhead is still too large and will consume significant channel resources.

Following the prior works, we consider a pair of encoder-decoder networks to compress and recover CSI, and a pair of quantizer-dequantizer for quantization. As shown in Fig. 2, $\bm{H} \in \mathbb{R}^{2\times N_t\times N_c}$ is fed into the encoder network, whose output is $M$ float parameters. Then, the float parameters are quantized to a bit stream by the quantizer, whose output is denoted by $\bm{c}$. The dequantizer and the decoder are applied at the BS. The dequantizer transforms $\bm{c}$ to float numbers and the decoder recovers the dequantizer output to $\bm{H} \in \mathbb{R}^{2\times N_t\times N_c}$. The compression ratio (CR) is defined as $\textrm{CR}\triangleq M/2N_tN_c$. The whole feedback procedure can be presented as
\begin{align} \label{e4}
&\bm{c} = \mathcal{Q}_e \left(f_e(\bm{H},\varTheta_e), \varTheta_{q}\right), \\
&\hat{\bm{H}} = f_d\left(\mathcal{Q}_d\left( \bm{c}, \varTheta_{dq} \right),\varTheta_d \right),
\end{align}
where $\bm{\hat{\bm{H}}}$ is the reconstructed and cropped CSI,  $f_e$ and $f_d$ denote the encoder and decoder networks, $\mathcal{Q}_e$ and $\mathcal{Q}_d$ denote the quantizer and dequantizer functions, $\varTheta_e$, $\varTheta_d$, $\varTheta_q$, $\varTheta_{dq}$ are the parameters of $f_e$, $f_d$, $\mathcal{Q}_e$ and $\mathcal{Q}_d$, respectively. The complete CSI can be obtained by zero-padding followed by inverse DFT operation on $\bm{\hat{\bm{H}}}$.

\subsection{Channel model}
\par Due to the large size of the antenna array implemented in massive MIMO systems, spherical wave channel model should be considered instead of a plane wave channel model \cite{5595728}. This is also verified through measurements in \cite{7062910} and \cite{6206345}. Spherical wave channel model is more realistic, and has been widely adopted in wireless communication applications \cite{8815472,8032474,9322174}. We also adopt a 3-D geometric stochastic channel model \cite{jaeckel2014quadriga} in this work, which incorporates the spherical wave channel model.
\begin{figure}[htbp]
\centering
\includegraphics[width=2.5in]{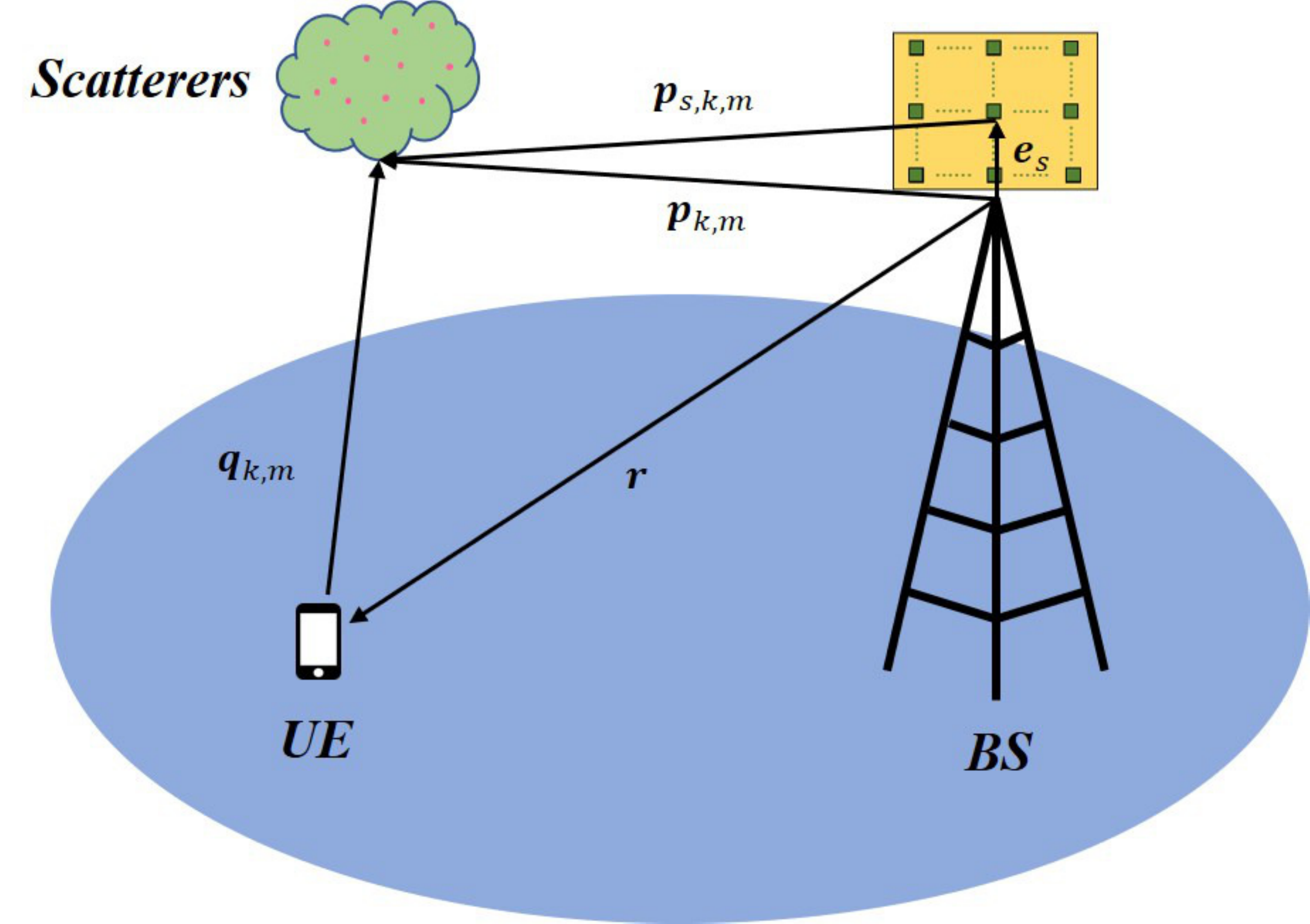}
\caption{The geometric relationship between the BS, UE and scatterers.}
\end{figure}
\par In particular, we model the differences of antennas and sub-paths of channel in detail by considering the scatterers\footnote{The following illustrations are based on the single-bounce model for simplifying notations, which can be easily extended to the multi-bounce model. In Section V, the proposed method is evaluated on the multi-bounce model. More details about the multi-bounce model can be found in \cite{jaeckel2014quadriga}.}. The initial length of the $k$-th path, $d_k$, can be expressed as
\begin{equation}
d_k = \Vert \bm{r} \Vert_2 + \tau_kc,
\end{equation}
where  $\|\bm{r}\|_2$ is the initial distance between the BS and the UE, $\tau_k$ is the delay of the $k$-th path, and $c$ is the speed of light. We define $\bm{q}_{k,m}$ as the arrival vector of the $m$-th sub-path in the $k$-th path pointing from the initial UE location to the scatterers. Then, its length can be expressed as
\begin{equation}
\Vert \bm{q}_{k,m} \Vert_2 = \frac{d_k^2 - \Vert \bm{r} \Vert_2^2}{2(d_k + \bm{r}^T\bar{\bm{q}}_{k,m})},
\end{equation}
where
\begin{equation}
\bar{\bm{q}}_{k,m} \triangleq
\begin{bmatrix}
\cos\phi_{k,m}^a\cos\theta_{k,m}^a \\
\sin\phi_{k,m}^a\cos\theta_{k,m}^a \\
\sin\theta_{k,m}^{a}
\end{bmatrix},
\end{equation}
$\phi_{k,m}^a$ and $\theta_{k,m}^a$ are the azimuth and elevation angle of arrival (AOA) of the $m$-th sub-path in the $k$-th path, respectively. Therefore, the location of the scatterers can be obtained by the geometric relationship shown in Fig. 3.

To model the spherical wave, each antenna element $s$ at the BS is considered separetly. Given an initial antenna element location and its departure vector of the $m$-th sub-path in the $k$-th path $\bm{p}_{k,m}$, the departure vector of the $s$-th antenna of the $m$-th sub-path in the $k$-th path $\bm{p}_{s,k,m}$, the corresponding elevation AOD $\theta^d_{s,k,m}$ and azimuth AOD $\phi^d_{s,k,m}$ can be derived as
\begin{equation}
\theta_{s,k,m}^d = \arcsin{\frac{p_{s,k,m,z}}{\Vert \bm{p}_{s,k,m} \Vert_2}}
\end{equation}
and
\begin{equation}
\phi_{s,k,m}^d = \arctan\frac{p_{s,k,m,y}}{p_{s,k,m,x}},
\end{equation}
where
\begin{equation}
\bm{p}_{s,k,m} = \bm{p}_{k,m} - \bm{e}_s,
\end{equation}
and $\bm{e}_s$ is the vector from initial antenna element to the $s$-th antenna element, while $p_{s,k,m,x}$, $p_{s,k,m,y}$ and $p_{s,k,m,z}$ are the Cartesian coordinate components of $\bm{p}_{s,k,m}$. Therefore, the deterministic phase $\psi_{s,k,m}$ and delay $\tau_{s,k}$ can be derived by
\begin{equation}
\psi_{s,k,m} = \frac{2\pi}{\lambda_c}(d_{s,k,m}\bmod\lambda_c)
\end{equation}
and
\begin{equation}
\tau_{s,k} = \frac{\sum_{m=1}^{M_k}d_{s,k,m}}{M_kc},
\end{equation}
where
\begin{equation}
d_{s,k,m} = \Vert \bm{p}_{s,k,m} \Vert_2 + \Vert \bm{q}_{k,m} \Vert_2,
\end{equation}
$M_k$ is the number of sub-paths in the $k$-th path, $\lambda_c$ is the wavelength and $\bmod$ stands for the module operation. The above modeling has taken spherical wave into consideration. Therefore, the channel between the $s$-th BS antenna and the UE via the $k$-th path can be described as (14),  which is shown on the top of the next page, where $\bm{P}_{s,k,m}$, $F_{\mathrm{rx}, \theta}$, $F_{\mathrm{rx}, \varphi}$, $F_{\mathrm{tx}, \theta}$, $F_{\mathrm{tx}, \varphi}$, $j$ and $\psi_{k,m}^{0}$ are the polarization coupling matrix of the $s$-th antenna of the $m$-th sub-path in the $k$-th path, elevation polarimetric antenna response at the receiver, azimuth polarimetric antenna response at the receiver, elevation polarimetric antenna response at the transmitter, azimuth polarimetric antenna response at the transmitter, imaginary unit and the random phase of the $m$-th sub-path in the $k$-th path, respectively.

\begin{figure*}[t]
\centering
    \begin{equation}
    g_{s,k} = \sum_{m=1}^{M_k}\left[\begin{array}{l}
F_{\mathrm{rx}, \theta}\left(\theta_{k,m}^{a}, \varphi_{k,m}^{a} \right) \\
F_{\mathrm{rx}, \varphi}\left(\theta_{k,m}^{a}, \varphi_{k,m}^{a} \right)
\end{array}\right]^{T}\bm{P}_{s,k,m}\left[\begin{array}{l}
F_{\mathrm{tx}, \theta}\left(\theta_{s,k,m}^{d}, \varphi_{s,k,m}^{d} \right) \\
F_{\mathrm{tx}, \varphi}\left(\theta_{s,k,m}^{d}, \varphi_{s,k,m}^{d} \right)
\end{array}\right]e^{\left(-j\psi_{k,m}^{0}-j\psi_{s,k,m}\right)}
    \end{equation}
\hrule
\end{figure*}
Therefore, the $(s,l)$-th element of $\tilde{\bm{H}}$ in spatial-frequency domain can be expressed as
\begin{equation}
\tilde{\bm{H}}_{s,l} = \sum_{k=1}^{K'}g_{s,k}e^{\left(-j2\pi\frac{l-1}{\tilde{N_c}}B\tau_{s,k}\right)},
\end{equation}
where $B$ is the bandwidth and $K'$ is the number of paths, $s=1,\cdots,N_t$ and $l=1,\cdots,\tilde{N_c}$.

\section{Problem Formulation}

In this section, we first formulate CSI feedback as a linear inverse problem following the related works. Then, the iterative shrinkage-thresholding algorithm (ISTA) will be introduced, which inspired the proposed method.

\subsection{Problem formulation}


We consider a learnable matrix as the encoder, which can be conveniently designed as a light linear layer, and is appropriate for the UE due to its limited computation and storage ability. Therefore, the projected vector $\bm{v}$ can be expressed as

\begin{equation}
\bm{v} = \bm{A}\bm{x},
\end{equation}
where $\bm{A}$ is the learnable matrix and $\bm{x} \in \mathbb{R}^{2N_tN_c}$ is the vector form of the CSI matrix $\bm{H}$. The decoder at the BS can be regarded as solving an inverse problem, which is presented as
\begin{equation}
\min_{\bm{x}} \frac{1}{2}\|\bm{v}-\bm{A}\bm{x}\|_2^2.
\end{equation}

Due to the huge dimension reduction, the problem (17) is highly ill-posed; and hence hard to solve directly. Typically, a regularization term is introduced into the optimization function to exploit any known prior information about the optimal solution. Therefore, the problem (17) can be modified as
\begin{equation}
\min_{\bm{x}} \frac{1}{2} \|\bm{v}-\bm{A}\bm{x}\|_2^2 + R(\bm{x}),
\end{equation}
where $R(\bm{x})$ is the regularization term.
\subsection{ISTA}

Considering the sparsity of CSI, conventional CS-based and model-driven DL methods utilize $l_1$-norm as the regularization term. Therefore, the problem (18) can be written as

\begin{equation}
\min_{\bm{x}} \frac{1}{2} \|\bm{v}-\bm{A}\bm{x}\|_2^2 + \lambda \|\bm{x}\|_1.
\end{equation}

ISTA \cite{FORNASIER2008187} is a classical iterative method to solve problem (19), and the related model-driven DL methods for CSI feedback \cite{wu2019compressed}, \cite{wang2020learnable} are inspired by it. Its iterative formulation at the $t$-th step can be expressed as
\begin{align}
&\bm{u}^{(t)} = \bm{x}^{(t-1)} - \alpha\bm{A}^{T}\left(\bm{A}\bm{x}^{(t-1)}-\bm{v}\right), \\
&\bm{x}^{(t)} = \mathrm{sign}(\bm{u}^{(t)})\mathrm{max}(\bm{0},|\bm{u}^{(t)}|-\bm{\theta}),
\end{align}
where $\bm{u}^{(t)}$, $\bm{0}$, $\bm{\theta}$ and $\alpha$ are the intermediate variable, zero vector, thresholding term and step size, respectively. The $\mathrm{sign}$ and $\mathrm{max}$ are element-wise operations, which can be expressed as
\begin{equation}
\mathrm{sign}(u) = \left\{ \begin{array}{lll}
1& \textrm{if $u>0$}\\
0& \textrm{if $u=0$}\\
-1& \textrm{otherwise}\\
\end{array}\right.
\end{equation}
and
\begin{equation}
\mathrm{max}(u,\theta) = \left\{ \begin{array}{ll}
u& \textrm{if $u \ge \theta$}\\
\theta & \textrm{if $u<\theta$}.\\
\end{array}\right.
\end{equation}

\section{Design of LORA and Training Scheme}


In this section, we first propose a novel model-driven DL method, called LORA, which unfolds the derived iterative formulations to a NN and incorporates a regularization learning module. Moreover, considering the quantization in CSI feedback procedure, the QAT and learnable quantization methods will be employed.
%

\subsection{Architecture of LORA}
\begin{figure*}[htbp]
\centering
\subfloat[]{
\includegraphics[width=2.4in]{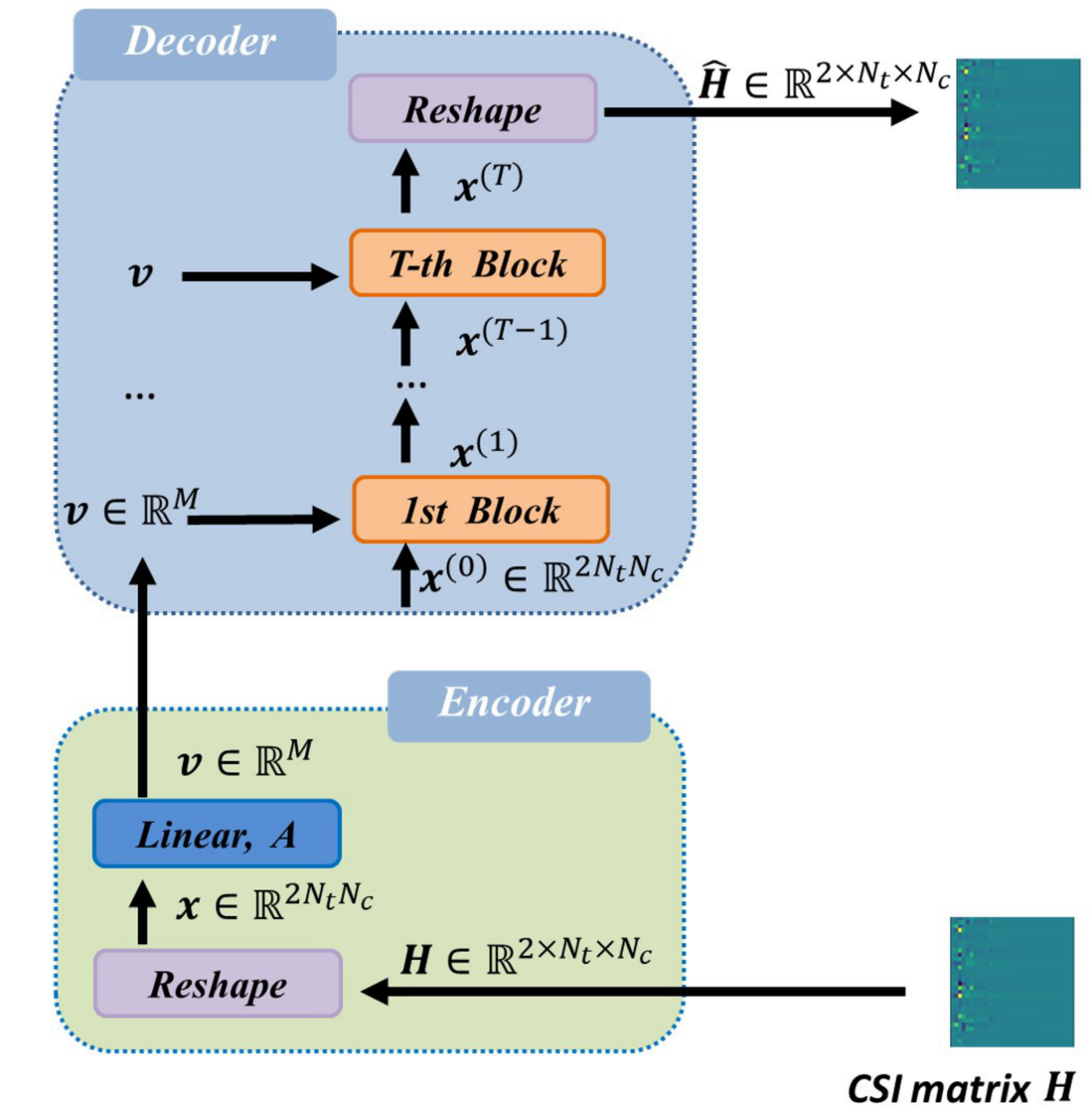}}
\hspace{+10mm}
\subfloat[]{
\includegraphics[width=2in]{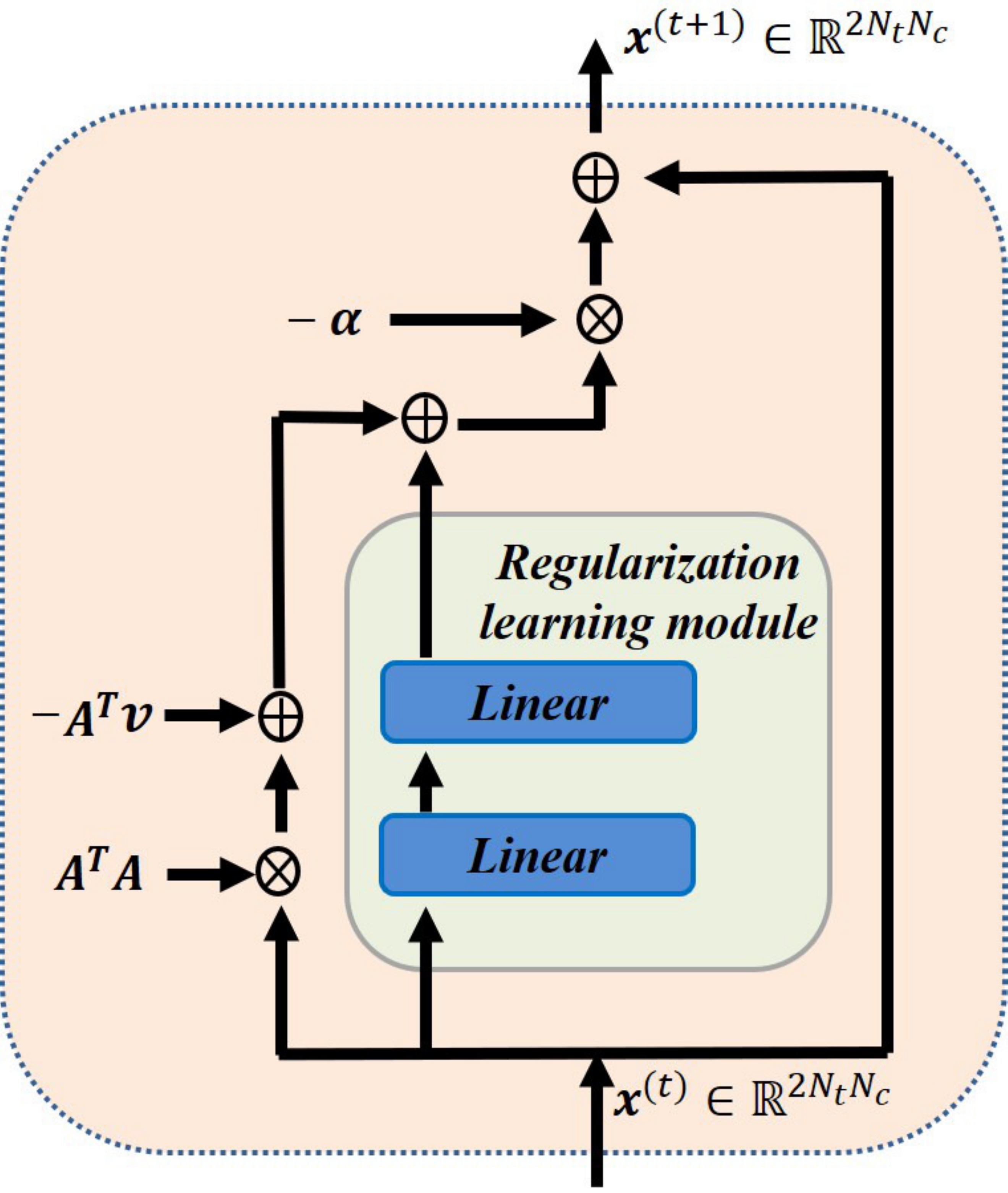}}
\caption{The NN architecture and forward procedure of LORA: (a) shows the overall architecture of LORA with the encoder and decoder. (b) shows the detailed architecture of one block in LORA, including the regularization learning module.}
\end{figure*}
As in Equation (16) presented above, we consider the CSI in the vector form. However, instead of fixing the regularization term to $l_1$-norm, we consider $R(\bm{x})$ as a learnable transform, which is assumed to be differentiable. Then, the iteration formulation at the $t$-th step ($t=1,\dots, T$) in the solution of problem (18) can be derived as
\begin{equation}
\bm{x}^{(t)} = \bm{x}^{(t-1)}-\alpha^{(t-1)}\left(\bm{A}^T(\bm{A}\bm{x}^{(t-1)}-\bm{v}) + \nabla R(\bm{x}^{(t-1)})\right),
\end{equation}
where $\nabla R(\cdot)$ stands for the gradient of $R(\cdot)$.
Therefore, each iteration can be designed as a block to develop the decoder of LORA. The step size $\alpha$ in (24) is also set as a learnable parameter that is different in each block. The matrix $\bm{A}$ is the same one used by the encoder. To learn $\nabla R(\cdot)$, each block employs a regularization learning module, which will be introduced in detail in the next sub-section. The overall architecture of LORA is presented in Fig. 4.

Model-driven DL exploits NNs to replace the explicit expressions or manually-set parameters in model-based methods. This can mitigate the performance loss due to inaccurate modeling, while exploiting the valuable knowledge of the model simultaneously. Besides, model-driven DL methods can also prevent the over-fitting problem, and are usually easier to train compared to purely data-driven NN approaches.

Notably, the initialization of $\bm{x}$ is important due to the use of gradient descent. Since CSI has sparsity, and its values are near zero, $\bm{x}^{(0)} = \bm{0}$ is considered as the initialization for LORA.

\subsection{Regularization learning module}


The parameters in ISTA, such as the measurement matrix, step size, etc, are all treated as learnable parameters in existing works. However, the regularization term is set as conventional $l_1$-norm. Meanwhile, $l_1$-norm is not a fully accurate prior because of the weak sparsity of CSI. Even with strict sparsity, the measurement matrix needs to satisfy RIP condition to ensure the exactly signal recovery by using $l_1$-norm. Therefore, making the regularization learnable to directly fit the characteristics of CSI is a promising approach to this problem.

\par The regularization learning module architecture is shown in Fig. 4, which is a light multi-layer perceptron (MLP). The number of neurons of the input layer, hidden layer, and output layer of the regularization learning module are set as $2N_tN_c$, 1024 and $2N_tN_c$, respectively. Therefore, the mathematical expression for the regularization learning module can be expressed as
\begin{equation}
\textrm{MLP}(\bm{\mathrm{x}}) = \bm{W_2}\sigma\left(\bm{W_1}\bm{\mathrm{x}}\right),
\end{equation}
where $\bm{W_1}\in \mathbb{R}^{1024 \times 2N_tN_c }$ and $\bm{W_2}\in \mathbb{R}^{2N_tN_c \times 1024 }$ are the parameters of the first and second linear layers, $\sigma(\cdot)$ is the rectified linear unit (ReLU) activation function and $\bm{\mathrm{x}}\in \mathbb{R}^{2N_tN_c}$ is the input of the regularization learning module.
\par The reasons for choosing MLP as the architecture of the regularization learning module are as follows: (1) MLP has been proved that it has universal approximation ability \cite{tolstikhin2021mlp}, which is suitable to characterize the complex properties of CSI; (2) The linear layer in MLP has a dense connection architecture, which can keep the original information as much as possible compared to a  local connection architecture, such as a convolution operator. Since the regularization term is the part of the optimization problem, which needs to be set down once the training is finished, the parameters of the MLP should be shared by all the blocks. Meanwhile, the number of trainable parameters will be reduced in this way.

%
%

%




\subsection{Training scheme with quantization}
LORA is trained in an end-to-end manner. Mean square error (MSE) is used as the loss function, which can be written as
\begin{equation}
L(\varTheta)=\frac{1}{\mathcal{T}}\sum_{i=1}^\mathcal{T}\|\hat{\bm{H}}_i-\bm{H}_i\|_2^2,
\end{equation}
where $\mathcal{T}$ is the total number of samples in the training set and $\varTheta$ is the parameter of NN. To avoid hyper-parameter tuning for dynamic learning rate adjustment operator, ADAM\cite{kingma2014adam} with fixed learning rate is applied as the optimization operator.

\par The quantization module is also considered in CSI feedback for practical implementation. The conventional quantization procedure can be generally summarized as

\begin{equation}
\mathsf{q} = \mathrm{round}\left(\mathrm{clip}\left(\frac{\mathsf{r}-\mathsf{z}}{\mathsf{s}}, \mathsf{n}, \mathsf{p} \right)\right),
\end{equation}
\begin{equation}
\mathsf{b} = \mathrm{num2bit}(\mathsf{q}),
\end{equation}
where $\mathsf{r}$, $\mathsf{s}$, $\mathsf{z}$ and $\mathsf{q}$ are the float number, scale, zero point vaule and integer number, $\mathsf{n}$ and $\mathsf{p}$ are the lower bound and upper bound of the clip function, $\mathrm{clip}$ and $\mathrm{round}$ are the cliping and rounding functions, $\mathrm{num2bit}$ is the function which converts an integer number to bits and $\mathsf{b}$ is the resultant bit stream.

The corresponding dequantization procedure can be expressed as
\begin{equation}
\bar{\mathsf{q}} = \mathrm{bit2num}(\mathsf{b}),
\end{equation}
\begin{equation}
\hat{\mathsf{r}} = \bar{\mathsf{q}} \times \mathsf{s} + \mathsf{z},
\end{equation}
where $\mathrm{bit2num}$ is the function which converts bits to integer vaules, $\bar{\mathsf{q}}$ is the integer vaules converted by $\mathrm{bit2num}$ function and $\hat{\mathsf{r}}$ is the dequantized float vaules corresponding to $\mathsf{r}$.
The quantization and dequantization operations can be regarded as two blocks, which are inserted to the end of the encoder and the begining of the decoder, respectively. Therefore, the forward procedure of NN can be regarded as CSI feedback with quantization and dequantization. Meanwhile, since end-to-end training is applied, the backward procedure can be regarded as learning parameters with the effect of quantization. However, the rounding function is not differentiable, which would prevent back-propagation during training. Instead, we can employ straight-through differentiation \cite{bengio2013estimating}, where we set
\begin{equation}
\frac{\partial \mathrm{round}(x)}{\partial x} \triangleq 1.
\end{equation}

The aforementioned training scheme is a modified QAT method, which is inspired by the QAT in NN quantization \cite{jacob2018quantization} and CsiNet+ \cite{8972904}. Although QAT achieves a reasonable performance, it still has the weakness that the scale and zero point value, which are essential parameters for quantization, are manually set in QAT. As the quantization procedure is embedded into the end-to-end training, the scale and zero point vaule can also be learned and trained jointly. Considering the back-propagation during training, the derivatives of scale and zero point value can be derived from (27), (30) and (31) as

\begin{equation}
\frac{\partial \hat{\mathsf{r}}}{\partial \mathsf{s}} \simeq \left\{ \begin{array}{ll}
- \frac{\mathsf{r}-\mathsf{z}}{\mathsf{s}}+\mathrm{round}(\frac{\mathsf{r}-\mathsf{z}}{\mathsf{s}})& \textrm{if $\mathsf{n}<\frac{\mathsf{r}-\mathsf{z}}{\mathsf{s}}<\mathsf{p}$}\\
\mathsf{n}\quad \textrm{or} \quad \mathsf{p} & \textrm{otherwise}\\

\end{array} \right.
\end{equation}
and
\begin{equation}
\frac{\partial \hat{\mathsf{r}}}{\partial \mathsf{z}} \simeq \left\{ \begin{array}{ll}
0& \textrm{if $\mathsf{n}<\frac{\mathsf{r}-\mathsf{z}}{\mathsf{s}}<\mathsf{p}$}\\
1& \textrm{otherwise}.\\

\end{array} \right.
\end{equation}
We name the scale learnable quantization method as LSQ, and both scale and zero point vaule learnable quantization method as LSZQ. The LSQ and LSZQ are modified versions of NN quantization methods in \cite{esser2019learned} and \cite{bhalgat2020lsq+}, respectively.

\section{Numerical Experiment}
In this section, we study the effect of different design options of LORA. Moreover, the numerical results evaluating the performance of LORA in terms of the reconstruction accuracy, channel estimation error and the complexity in comparison with other baselines are presented.

\subsection{Experiment settings}

\subsubsection{Data generation}
QuaDRiGa \cite{jaeckel2014quadriga} is a general channel simulator that meets the 3GPP standards.
Moreover, the spherical waves introduced in Section II as well as other realistic scenarios can be modeled by QuaDRiGa. Therefore, in this work, we use QuaDRiGa to generate CSI matrices in rural macro non-line-of-sight (RMANLOS), urban macro non-line-of-sight (UMANLOS), and urban micro non-line-of-sight (UMINLOS) scenarios. The carrier frequency, number of subcarriers, subcarrier interval and $N_c$ are set as 3.5GHz, 1024, 30kHz and 32 for the above three scenarios. The BS is equipped with a cross-polarized uniform planar array (UPA) with half wavelength antenna spacing and $N_t = 32$ antennas. The UE is assumed to move along a linear trajectory with a velocity of $\hat{v}=$ 6km/h. The heights of the BS are 10m, 10m, and 25m for RMANLOS, UMINLOS and UMANLOS, respectively. Training and test datasets are generated with 40000 and 10000 samples, respectively.

\subsubsection{Training settings and evaluation metric}

LORA is implemented in PyTorch. The parameters of the NN are updated and optimized by the ADAM optimizer with default settings. The learning rate, number of epochs and batch size are set to 0.001, 1000 and 200, respectively. We use the normalized mean square error (NMSE) as the evaluation metric, which is defined as
\begin{equation}
\textrm{NMSE} = \mathbb{E}\left\{\frac{\|\bm{H}-\hat{\bm{H}}\|_2^2}{\|\bm{H}\|_2^2}\right\}.
\end{equation}
Unless stated otherwise, all experiments are implemented with the above settings.

\subsection{Ablation studies for the LORA architecture}

In this sub-section, the effects of different design options of LORA will be studied through ablation. The design motivations presented in Section IV are supported by the following numerical results.

\begin{figure}[t]
\centering
\subfloat[]{
\includegraphics[width=0.3\columnwidth]{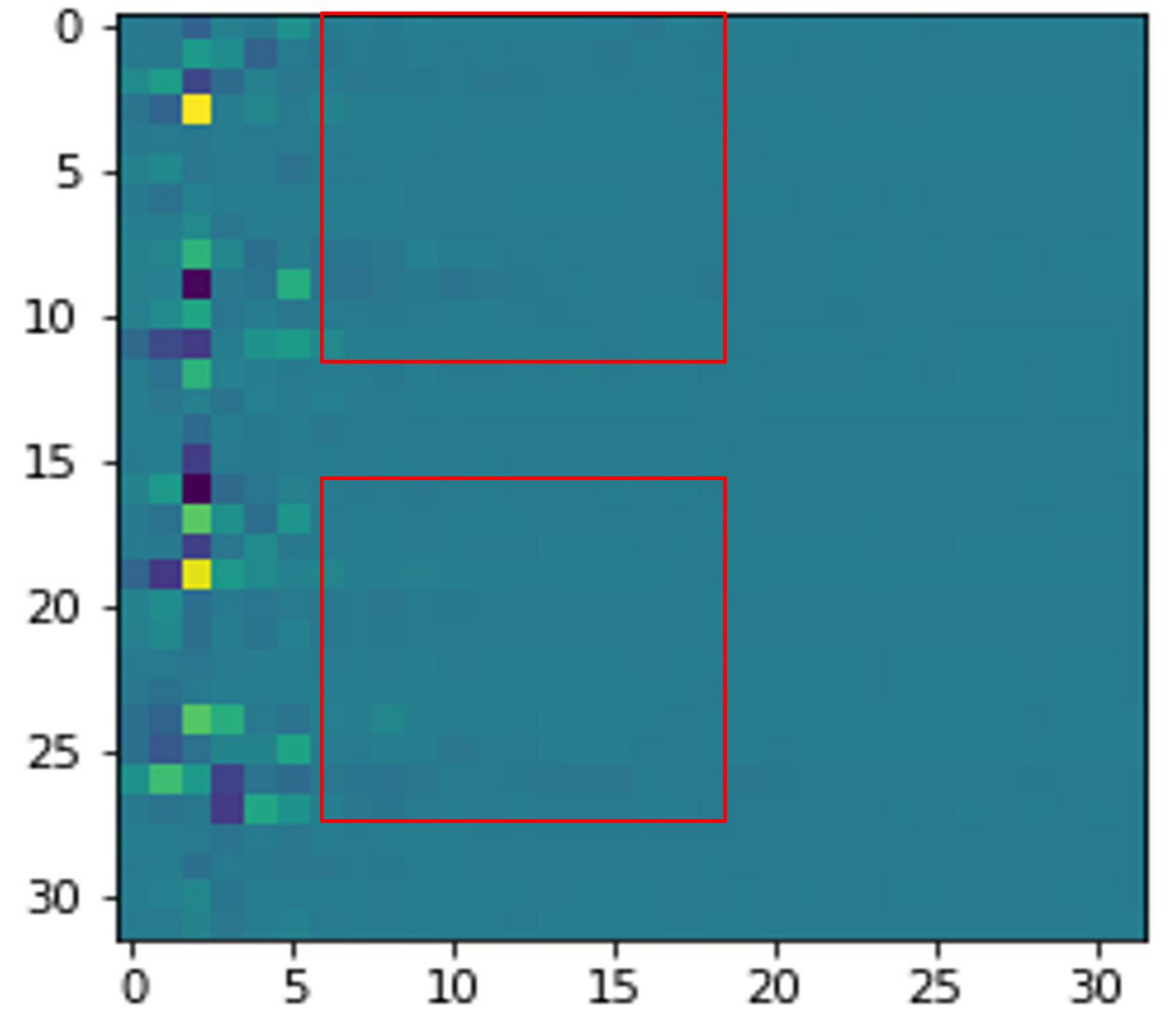}}
\subfloat[]{
\includegraphics[width=0.3\columnwidth]{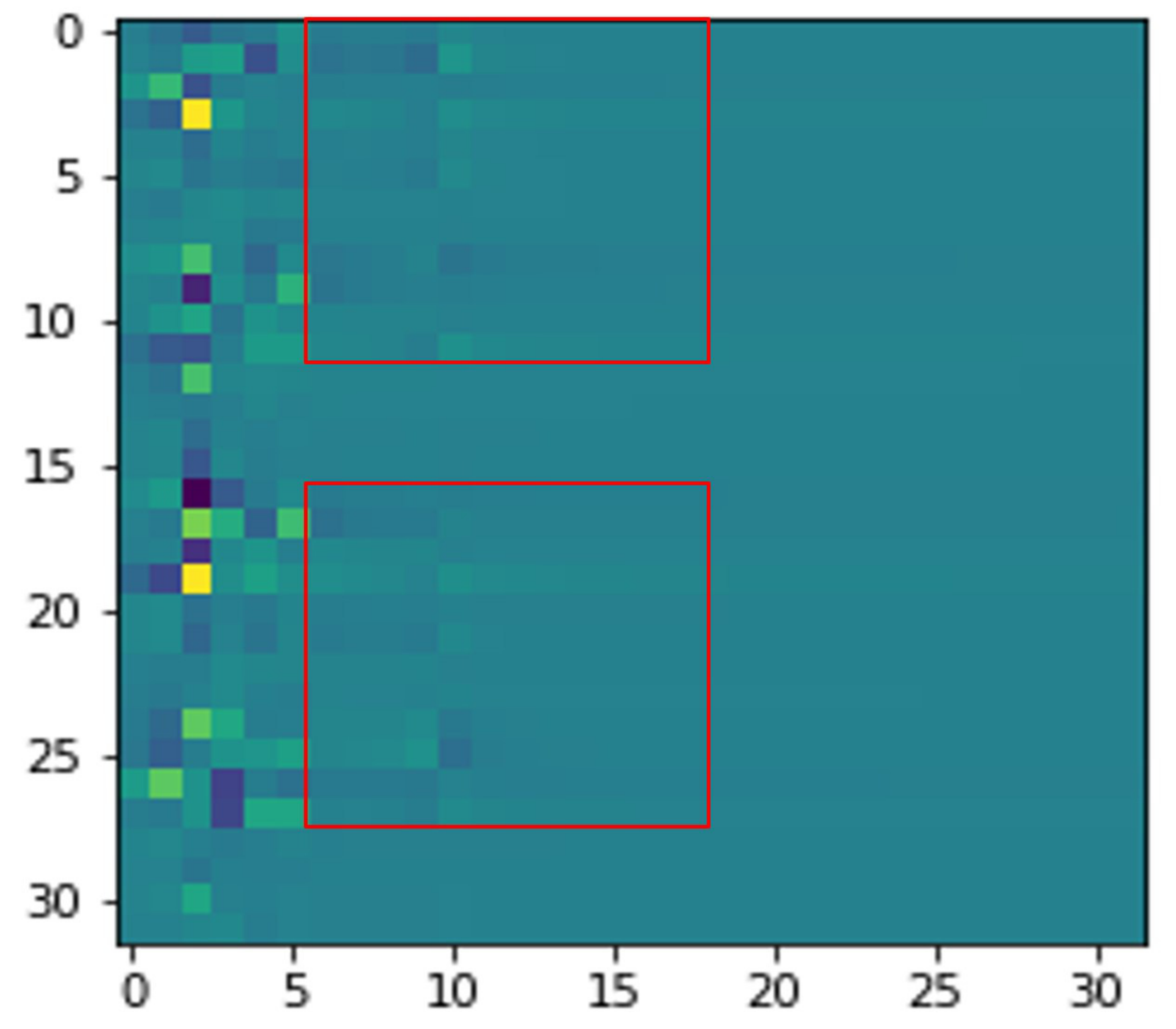}}
\subfloat[]{
\includegraphics[width=0.3\columnwidth]{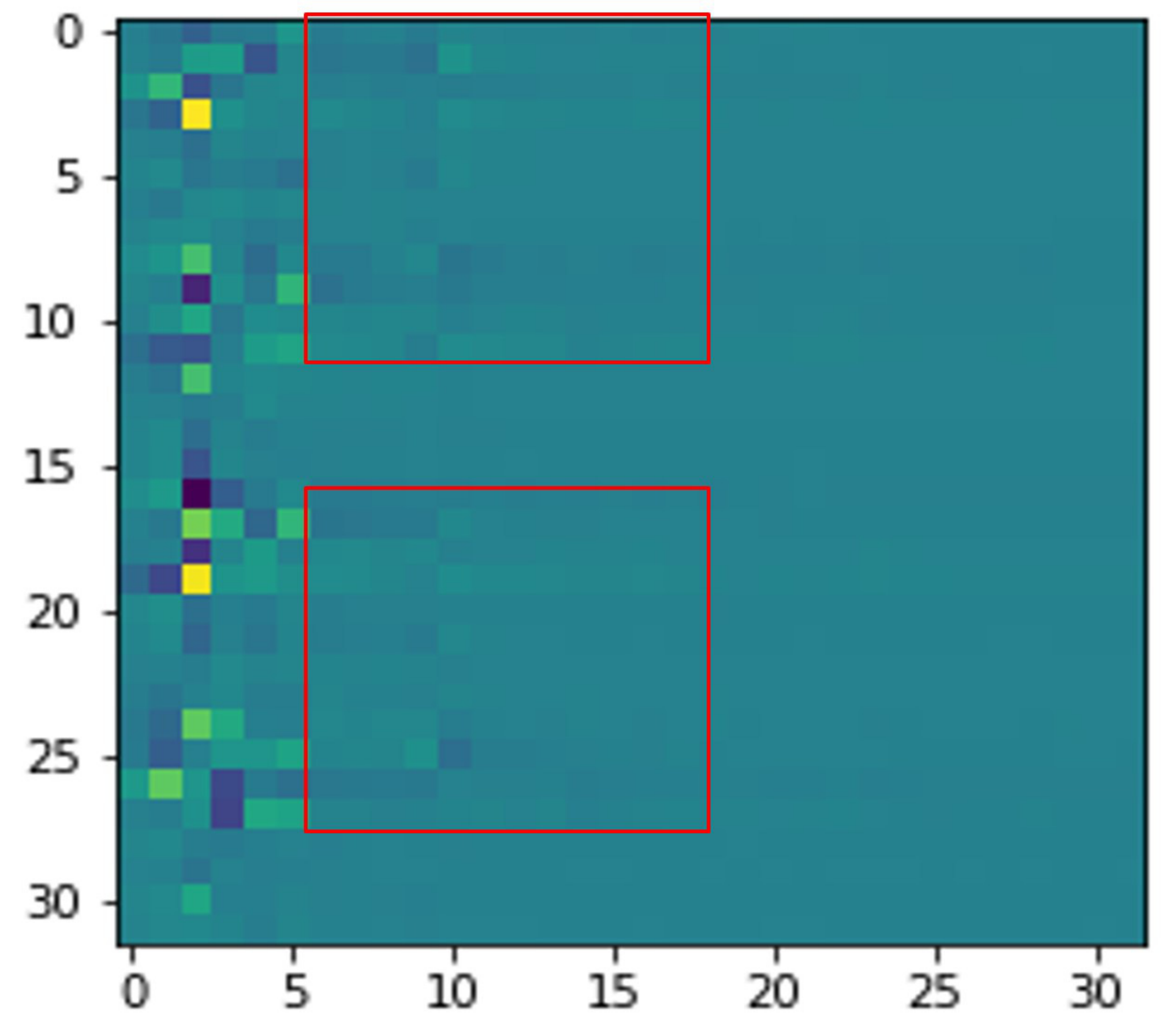}}
\caption{The visualization results of the last block of a trained LORA using MLP and CNN as two alternative architectures for the regularization learning module. (a) is the result of using CNN, while (b) is the result of using MLP. (c) is the groundtruth. }
\end{figure}

\subsubsection{Architecture of the regularization learning module}

\par To further illustrate the reasons of using MLP as the regularization learning module, we present the output of the last block using MLP and CNN\footnote{The CNN here consists of two convolutional layers with ReLU activation function. The convolution filter sizes are $3 \times 3 $ and the number of convolution channels is 32 and 2, respectively. } in Fig. 5. The part of the output of the last block, which corresponds to the real part of the CSI, are shown in the figures. By comparing the visualization results, it can be seen that using CNN loses some information in the region of red box, while the MLP recovers them well. Numerical results are also presented in TABLE I to compare the NMSE performance achieved by the two architectures. The above visualization and numerical results show that using MLP for the regularization learning module in LORA achieves a better performance than using CNN.

\begin{table}[t]
\centering
\caption{The NMSE performance of the output of each block of a trained LORA using MLP and CNN under CR$=1/16$ in the RMANLOS scenario. }
\begin{tabular}{|c|c|c|}
\hline
 \textbf{Block order}&\textbf{MLP}&\textbf{CNN}\\
\hline
1&\textbf{0.2936}&0.317\\
\hline
2&\textbf{0.0888}&0.5911\\
\hline
3&\textbf{0.0174}&0.0848\\
\hline
4&\textbf{0.0012}&0.0753\\
\hline
\end{tabular}
\end{table}


\subsubsection{Effect of learnable part}
\begin{figure}[t]
\centering
\includegraphics[width=3in]{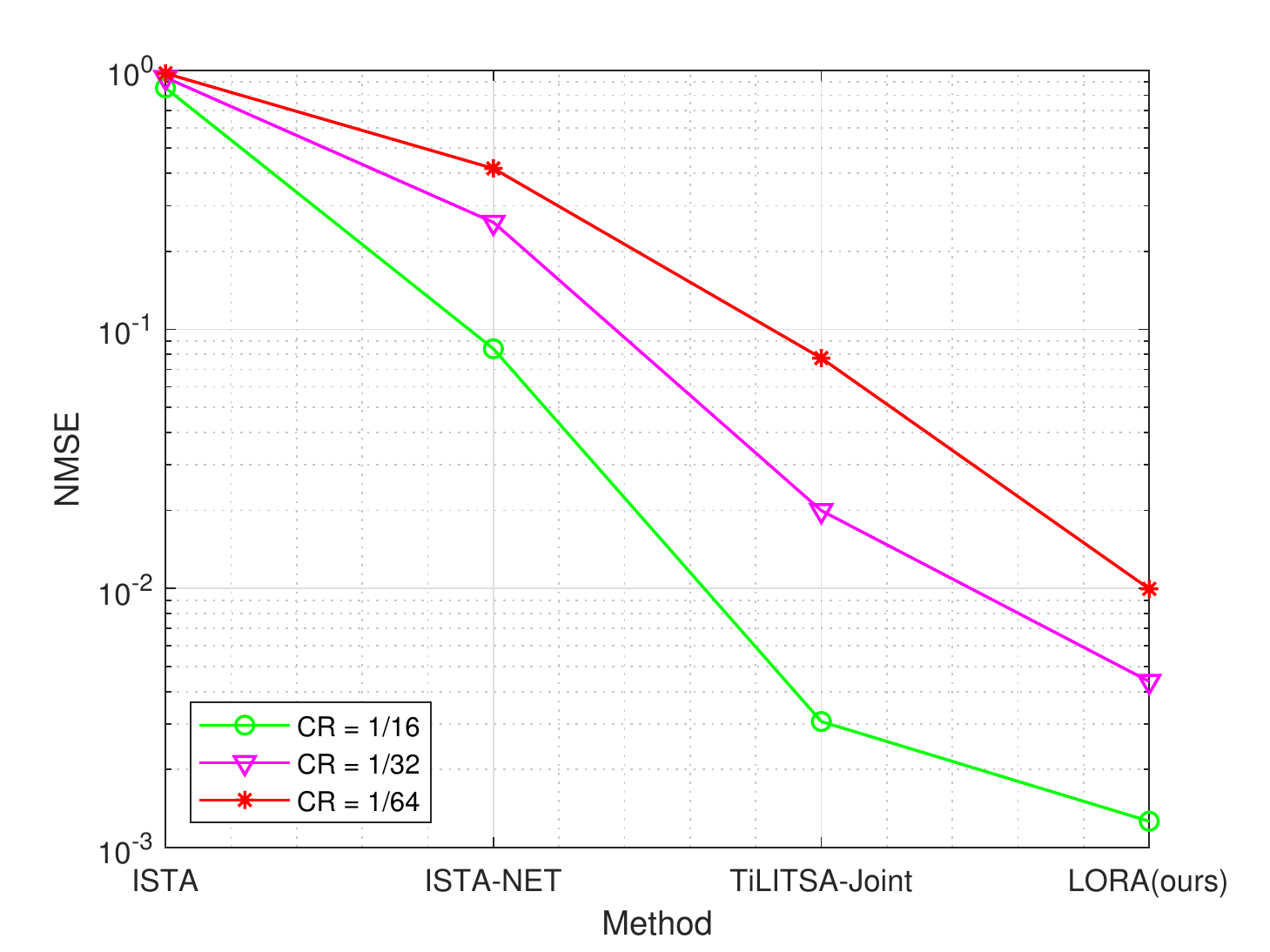}
\caption{The NMSE performance of different ISTA-based methods under different CR values in the RMANLOS scenario.}
\end{figure}

To verify the effectiveness of the regularization learning module in LORA, and to motivate the particular architecture argued for in Section IV-A, the performance of conventional ISTA\footnote{We choose Gaussian matrix as the measurement matrix.}, ISTA-NET, TiLISTA-Joint and LORA are compared next. The ISTA-NET stands for an ISTA unfolding method, which has learnable measurement matrix, step size and threshold. In TiLISTA-Joint, in addition to the parameters in ISTA-NET, a sparse transformation is also learned. In this experiment, the conventional ISTA is the baseline method, which has no learnable part. The results are shown in Fig. 6. As we can see, the performance increases as we learn more parameters of the underlying model. This phenomenon suggests that the performance can be improved by making more of the model parameters learnable. Specifically, the results verify that the learnable regularization outperforms $l_1$-norm regularization term. In addition, the importance of regularization in terms of the performance can be shown among the compared DL methods.

We also compare the $l_1$-norm of the output of the ISTA-NET and LORA, to further study the learned regularization term. The results are shown in TABLE II. It can be seen that the $l_1$-norm of the output has obvious difference between ISTA-NET and LORA in different scenarios. Although the learned regularization term has no explicit formulation, we can conclude that the learned regularization term is different with the conventional $l_1$-norm according to the NMSE performance and the $l_1$-norm of the output.

\begin{table}[t]
\centering
\caption{The $l_1$-norm of the output of different methods under CR$=1/16$ in the UMANLOS and UMINLOS scenario.}
\begin{tabular}{|c|c|c|}
\hline
\diagbox{\textbf{Methods}}{\textbf{Scenarios}}&UMANLOS&UMINLOS\\
\hline
ISTA-NET&3.74&8.18\\
\hline
LORA(ours)&4.39&10.15\\
\hline
\end{tabular}
\end{table}

\subsubsection{Effect of different blocks of LORA and the width of MLP}
\begin{table}[t]
\centering
\caption{The NMSE performance of LORA with different number of blocks and width of MLP under CR$=1/64$ in the RMANLOS scenario.}
\begin{tabular}{|c|c|c|c|c|}
\hline
\diagbox{\textbf{Neurons}}{\textbf{Blocks}}&4&5&6&7\\
\hline
512&0.1791&0.1706&0.1656&0.1625\\
\hline
1024&0.0099&0.0071&0.0055&0.0047\\
\hline
4096&0.1629&0.0137&0.0105&0.0081\\
\hline
\end{tabular}
\end{table}

The number of blocks of LORA and the neurons of hidden layer in MLP are investigated. The impact of these two hyper-parameters on the performance of LORA is shown in TABLE III. If the performance is more important than complexity in some scenarios, such as the high accuracy communication, LORA has the potential to improve the performance by using more blocks according to the results in TABLE III. However, it is worth to mention that increasing the width or depth of a NN may cause the difficulties in training, such as gradient vanishing and explosion. The training parameters, such as the learning rate and batch size, need to be carefully adjusted.


\subsection{Performance results}
\begin{figure}[t]
\centering
\includegraphics[width=3in]{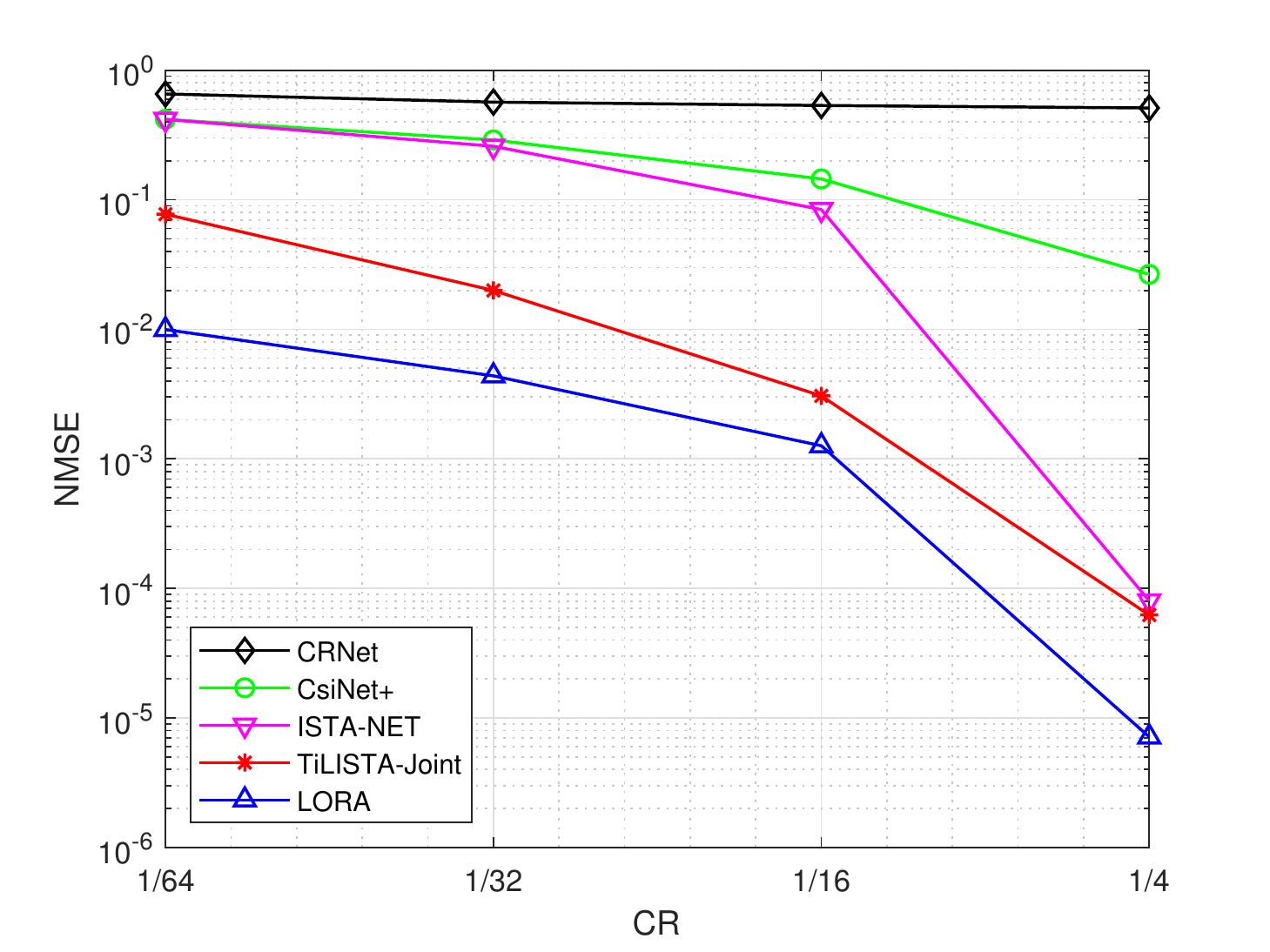}
\caption{The NMSE performance versus CR for different methods in the RMANLOS scenario.}
\end{figure}
\begin{figure}[t]
\centering
\includegraphics[width=3in]{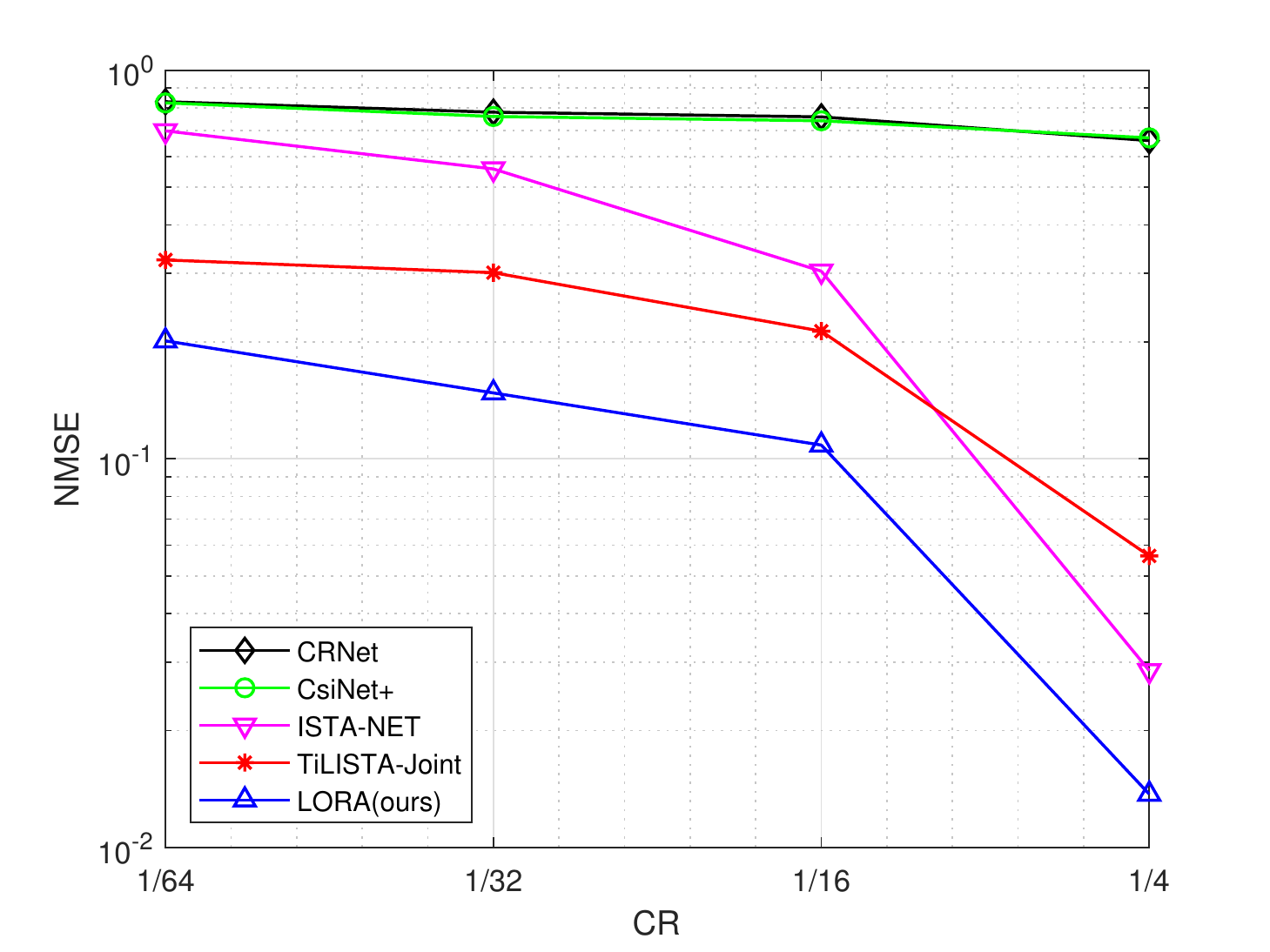}
\caption{The NMSE performance versus CR for different methods in the UMANLOS scenario.}
\end{figure}

The performance of LORA is evaluated in the aforementioned three scenarios, and compared with four DL benchmarks to investigate the effectiveness and robustness of LORA. CsiNet+ and CRNet are considered as two CNN-based benchmarks. Meanwhile, TiLISTA-Joint and the ISTA-NET are considered as the unfolding based benchmarks. The results are shown in Fig. 7 - Fig. 9 for the three scenarios RMANLOS, UMANLOS and UMINLOS, respectively. LORA outperforms all four benchmarks clearly under all CR values in all the scenarios, especially under small CR values. The presented results demonstrate the superiority of LORA compared with two CNN-based and two unfolding based methods in terms of the recovery accuracy even under $\textrm{CR}=1/64$. Moreover, the robustness of LORA in terms of achieving a superior performance in a variety of communication scenarios and CR values are verified. Therefore, LORA is a promising method in practice with high performance and small overhead.

\begin{figure}[t]
\centering
\includegraphics[width=3in]{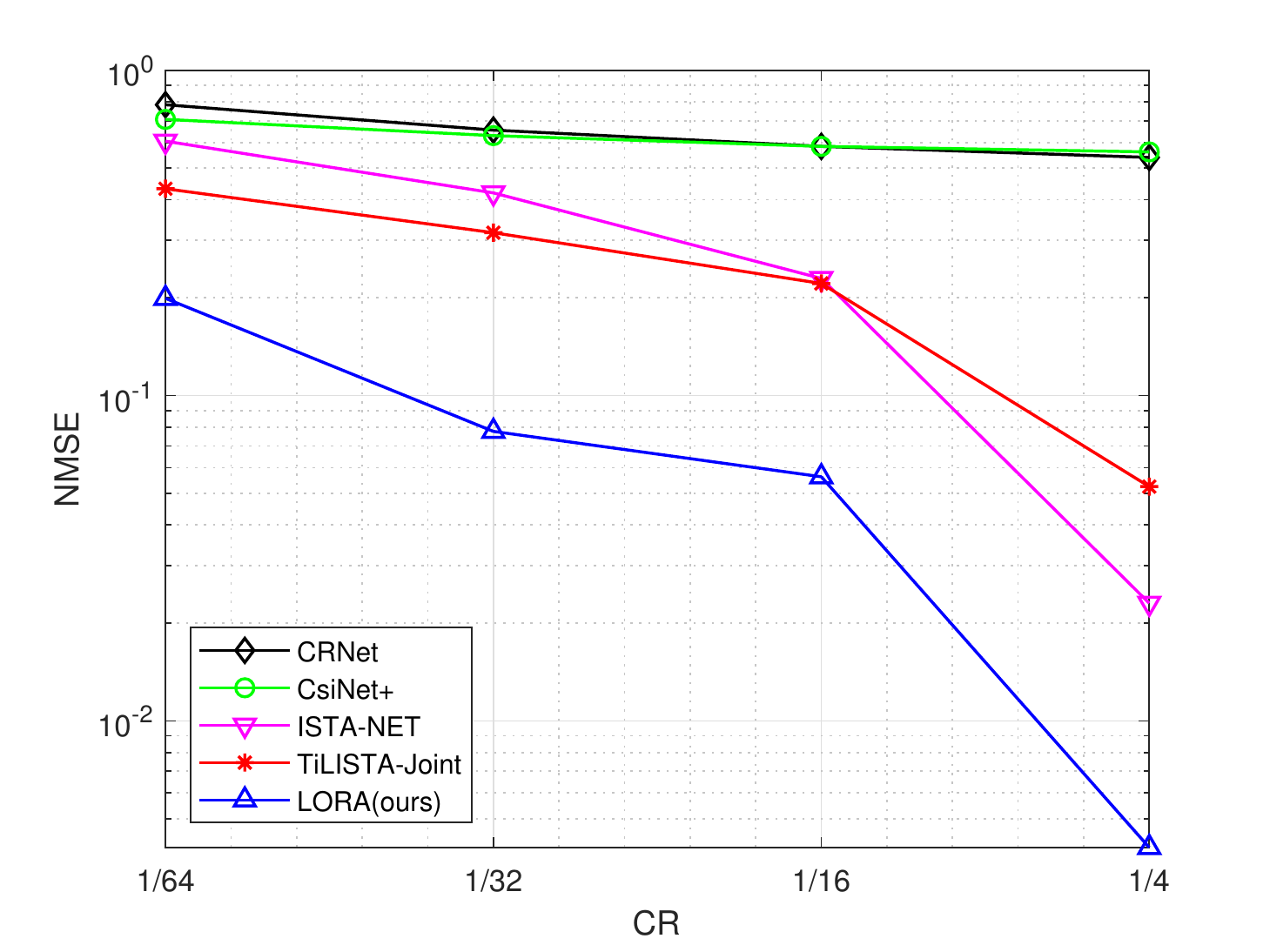}
\caption{The NMSE performance versus CR for different methods in the UMINLOS scenario.}
\end{figure}

Next, we study the impact of channel estimation errors on the performance of LORA. In particular, different from the above experiments that assumed perfect CSI at the UE, the additive white Gaussian noise is added to the CSI as the input during both training and testing stages. The NMSE performance under CR$=1/64$ of TiLISTA-Joint and LORA in the RMANLOS scenario are compared. The signal-to-noise ratio (SNR) is used to adjust the noise level introduced to the CSI. According to the results in Fig. 10, LORA significantly outperforms TiLISTA-Joint in all noise levels, which suggests that LORA has better capability to adapt to different degrees of channel estimation errors. Notably, the NMSE here is defined as
\begin{equation}
\textrm{NMSE} = \mathbb{E}\left\{\frac{\|\bm{H}_{gt}-\hat{\bm{H}}\|_2^2}{\|\bm{H}_{gt}\|_2^2}\right\},
\end{equation}
where $\bm{H}_{gt}$ denotes the perfect CSI. Since the input of LORA contains channel estimation errors, the results in Fig. 10 also show the denoising capability of LORA in addition to the CSI compression and recovery.
\begin{figure}[t]
\centering
\includegraphics[width=3in]{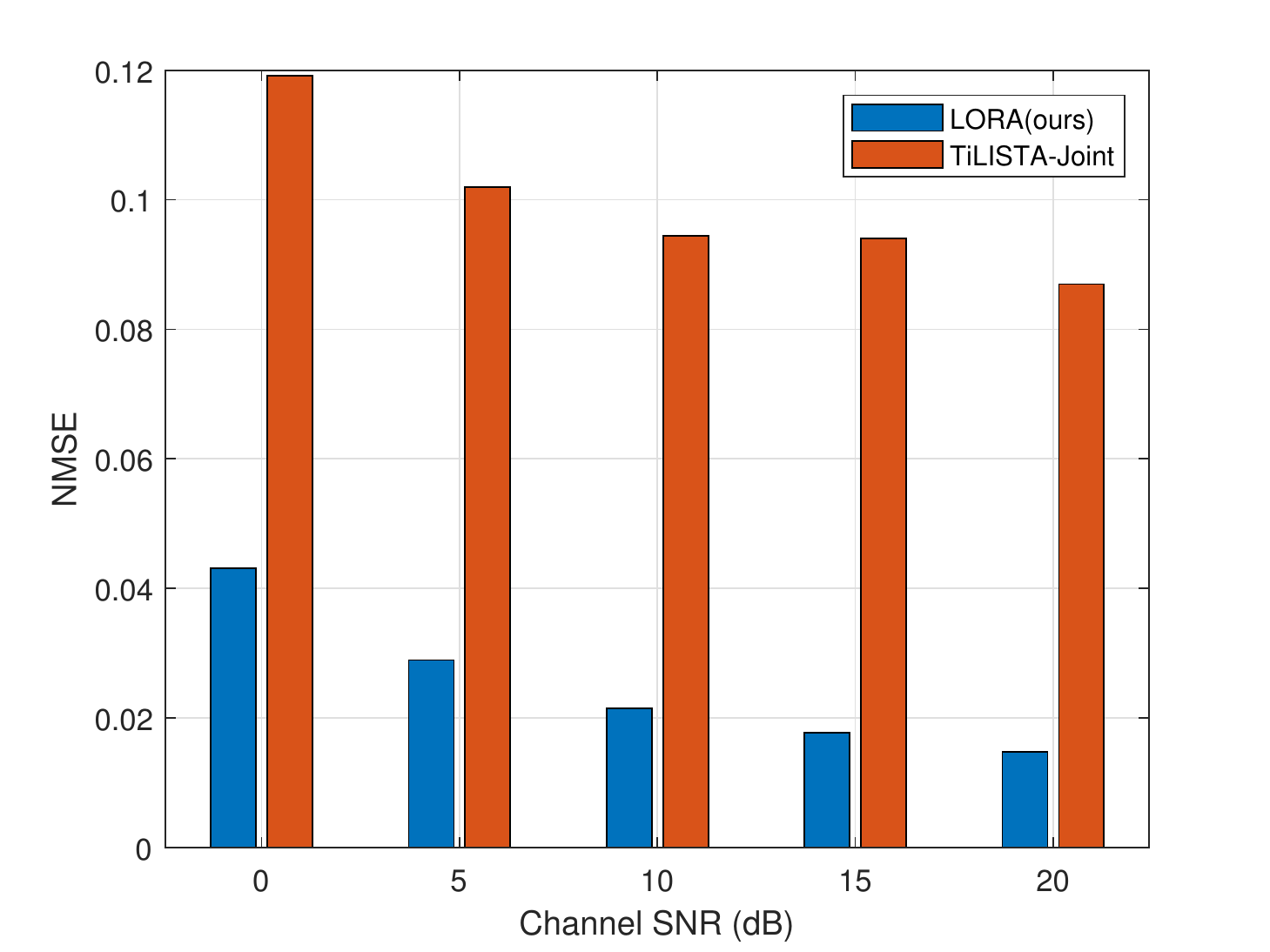}
\caption{The NMSE performance with different SNRs.}
\end{figure}

\subsection{Complexity}
Here, we analyze the storage and computational complexity of LORA. At the encoder of LORA, there is only a measurement matrix, which has $\left((2N_tN_c)^2 \times \textrm{CR}\right)$ parameters. At the decoder of LORA, each block has a learning rate parameter and a regularization learning module, which has $(4N_tN_c\times1024)$ parameters. Thus, the total number of parameters of LORA is $\left((2M+128^2)N_tN_c+4\right)$. However, due to the parameters sharing of regularization learning module for each block, the number of trainable parameters of LORA is $\left((2M+4096)N_tN_c+4\right)$.
\begin{table}[t]
\centering
\caption{The number of parameters of different methods in the encoder and decoder, respectively.}
\label{table 1}
\begin{tabular}{|c|c|c|c|c|}
\hline
\multirow{2}{*}{\textbf{Methods}}&\multicolumn{2}{|c|}{CR=1/4}&\multicolumn{2}{|c|}{CR=1/64}\\
\cline{2-5}
&\textbf{Encoder}&\textbf{Decoder}&\textbf{Encoder}&\textbf{Decoder}\\
\hline
CsiNet+ &1,048,772 &1,069,936&65,732 &86,896\\
\hline
TiLISTA-Joint &1,048,576 &11,468,820&65,536&10,485,780\\
\hline
LORA(ours)&1,048,576 &16,777,220&65,536&16,777,220\\
\hline
\end{tabular}
\end{table}
The number of parameters of CsiNet+, TiLISTA-Joint and LORA at the encoder and decoder are also shown in TABLE IV. We also consider CR$=1/4$ and CR$=1/64$ to represent the large CR and small CR cases. The results show that unfolding-based methods have less parameters at the encoder, but more parameters are needed at the decoder. It is because the convolution operator shares parameters and has less connections with the output of the former layer than MLP. However, since the BS usually has large storage, it is desirable to employ a larger model on the decoder side.

The main operator employed in LORA is a linear layer, whose computational cost can be calculated as $N_{in} \times N_{out}$. $N_{in}$ and $N_{out}$ are the sizes of input and output of the linear layer, respectively. Therefore, the encoder, decoder and total computational complexity of LORA are $\mathcal{O}(MN_cN_t)$, $\mathcal{O}(MN_t^2N_c^2)$ and $\mathcal{O}(MN_t^2N_c^2)$, which indicates that the computational cost of LORA depends on the number of antennas, preserved sub-carriers and feedback parameters.
\begin{table}[t]
\centering
\caption{The computation cost(s) of different methods with CR$=1/64$ in RMANLOS scenario.}
\begin{tabular}{|c|c|c|}
\hline
\textbf{Methods}&Train&Test\\
\hline
ISTA &-&90.96\\
\hline
CsiNet+ &64.307 &6.712\\
\hline
TiLISTA-Joint &109.688&10.432\\
\hline
LORA(ours)&\textbf{29.337} &\textbf{3.54}\\
\hline
\end{tabular}
\end{table}
In following, CR$=1/64$ is used as an example to compare the running time of one epoch of three DL based methods, and the results are shown in TABLE V. The key hardwares used in this experiment are i5-9400F CPU and RTX2080 GPU. According to TABLE V, it is obvious that LORA costs the least time among the compared methods, which shows the superiority of LORA in terms of computing time. Moreover, by comparing ISTA and learnable ISTA methods (i.e., TiLISTA-Joint and LORA), it can be inferred that introducing learnable regularization term improves the recovery performance while reducing inference time simultaneously. By separating the training and testing parts, it can be clearly observed that DL based methods have faster inference speed. Although the proposed method will cost extra time in training compared with conventional ISTA, online fine-tuning or MAML \cite{finn2017model} can be exploited to highly reduce the number of training epochs.

\begin{figure}[t]
\centering
\includegraphics[width=3.2in]{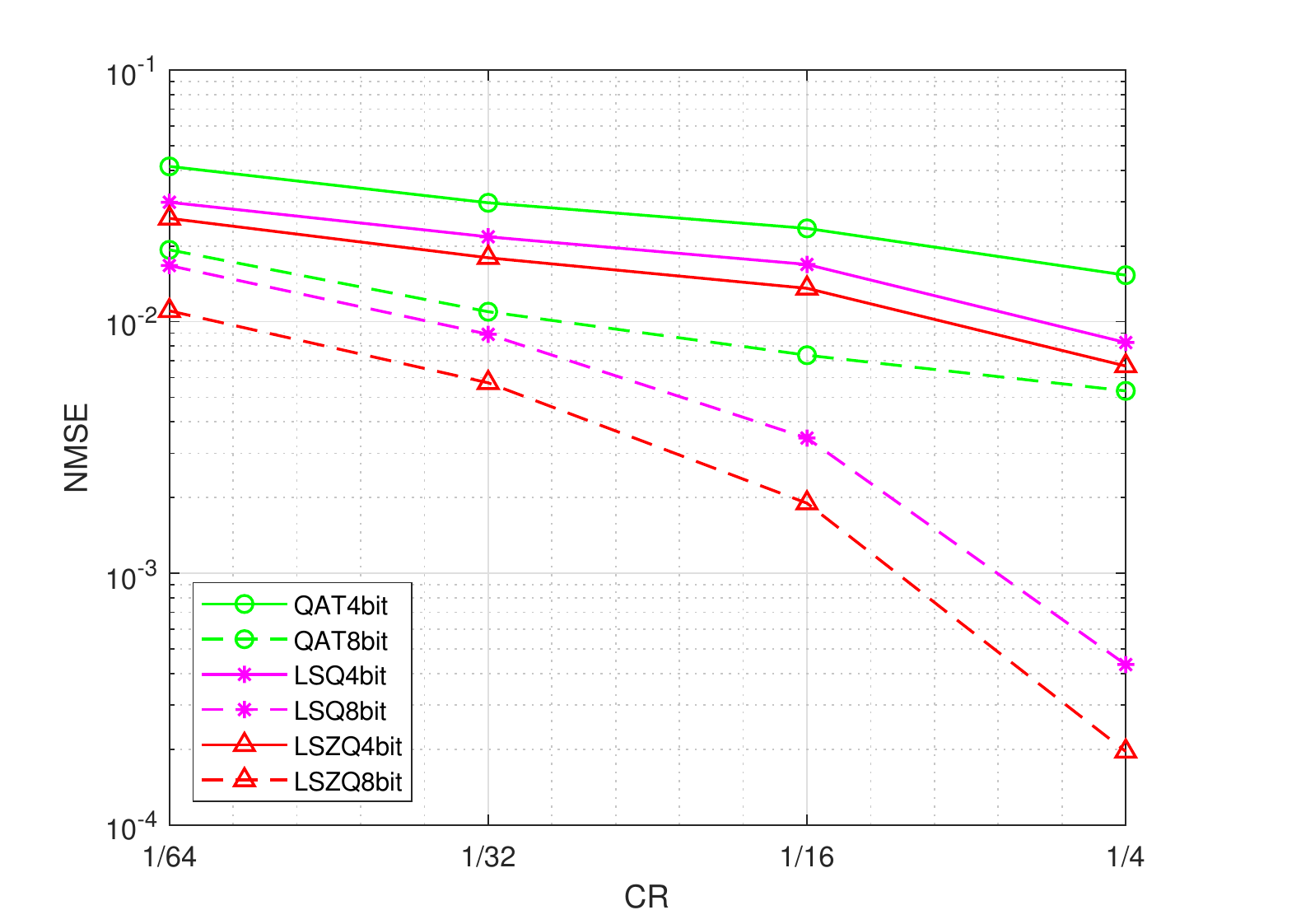}
\caption{The NMSE performance versus CR for different quantization methods and bit levels in the RMANLOS scenario.}
\end{figure}
\subsection{Quantization}

Different from the above simulations that ignore the quantization error, the proposed quantization method is evaluated in this sub-section. We first compare joint training LORA and the quantizer-dequantizer, and training the quantizer-dequantizer on a well-trained LORA. The NMSE of these two approaches are 0.0053 and 0.0069 under CR$=1/4$ in the RMANLOS scenario. Hence, as one would expect, joint training is preferable as the NN parameters adapt to the quantization effect.

We next compare three methods with four different CRs and two different bit levels in the RMANLOS scenario. The NMSE performance of the quantization methods are presented in Fig. 11. QAT is the baseline method for comparison. From the figure, it is seen that LSQ and LSZQ both outperform QAT. Moreover, the performance increases with the increase in the number of  learnable parameters in the quantization module; that is LSZQ outperforms LSQ. The results not only verify that the scale and zero point value are essential parameters of the quantization, but also demonstrate the effectiveness of making these parameters learnable. As expected, the performance of all three methods improve with the number of quantization bits. In addition, the performance of LSQ and LSZQ increase with the increase in the CR more significantly in 8-bit quantization than 4-bit quantization.  It is also interesting that the improvement of LSQ and LSZQ compared to QAT also increases with the number of quantization bits. This phenomenon may be attributed to that the impacts of scale and zero point value on quantization are amplified due to the increase in the number of quantization bits.

\section{conclusion}

In this paper, a model-driven DL method, called LORA, has been proposed for efficient CSI feedback in FDD massive MIMO systems. LORA is constructed by unfolding an iterative optimization algorithm with learnable parameters. In particular, the derivatives of the regularization term of the optimization problem is parameterized as a learnable MLP to automatically and directly extract the characteristics of CSI rather than using the fixed conventional $l_1$-norm. Besides, a scale and zero point value learnable quantization method with end-to-end training was proposed to ease the performance decay caused by quantization. The numerical results not only show the effects of the various components of LORA, supporting the presented architecture, but also demonstrate the superiority and robustness of this architecture with respect to the existing techniques in the literature. It has been also shown that LORA with the proposed quantization method can be effective at different bit levels providing, flexibility in terms the available feedback channel capacity.

\ifCLASSOPTIONcaptionsoff
  \newpage
\fi

\bibliographystyle{IEEEtran}
\bibliography{ref}

\end{document}